\def\upd{{\rm d}}

\documentclass[prl,twocolumn,showpacs]{revtex4}
\usepackage{graphicx}

\begin{document}

\title{Fluctuation properties of an effective nonlinear system subject to Poisson noise}

\author{A. Baule and E. G. D. Cohen}

\affiliation{                    
The Rockefeller University, 1230 York Avenue, New York, NY 10065, USA
}

\begin{abstract}

We study the work fluctuations of a particle, confined to a moving harmonic potential, under the influence of friction and external Poissonian shot noise. The asymmetry of the noise induces an effective nonlinearity in the potential, which in turn leads to singular features in the work distribution. On the basis of an analytic solution we find that the conventional fluctuation theorem is violated in this model, even though the distribution exhibits a large deviation form. Furthermore, we demonstrate that the interplay of the various time scales leads to critical behaviors, such as a negative fluctuation function and a divergence in the work distribution at the singularity. In a certain parameter regime large negative work fluctuations are more likely to occur than the corresponding positive ones, though the average work is always positive, in agreement with the second law.

\end{abstract}

\pacs{05.40.-a, 05.70.Ln, 02.50.-r}

\date{\today}

\maketitle

Fluctuation theorems play a prominent role in the study of nonequilibrium fluctuations and have been derived for a variety of different systems and thermostatting mechanisms \cite{EvansD93,Gallavotti95,Kurchan98,Lebowitz99}. In a nonequilibrium steady state, which represents the next simplest generalization of equilibrium, the so-called \textit{conventional fluctuation theorem} states that the probability distribution $\Pi(a_\tau)$ of finding a particular value of a thermodynamic quantity $a_\tau$ (usually work or heat) over time $\tau$ satisfies a certain symmetry relation, which, for stochastic systems, can be formulated as \cite{Kurchan98,Lebowitz99}
\begin{eqnarray}
\label{conventional}
\frac{\Pi(a_\tau)}{\Pi(-a_\tau)}\cong e^{\beta a_\tau},
\end{eqnarray}
where $\cong$ indicates the behavior for large $\tau$. Eq.~(\ref{conventional}) is a refinement of the second law, in that it quantifies the probability of observing temporary second law violations (negative $a_\tau$) in the steady state. The validity of Eq.~(\ref{conventional}) is commonly related to the existence of a large deviation form of the probability distribution \cite{Lebowitz99,Derrida07}.
In the present letter we further investigate the relationship between large deviations and the fluctuation theorem. We show, by explicitly calculating the work distribution in an effectively non-linear Markovian Langevin model, that the conventional fluctuation can be violated even though the distribution admits a large deviation form. Our model exhibits a range of other striking properties such as singularities and critical behavior. These features are all explained on the basis of the time scales in the model, as we discuss in detail in the following

We consider an overdamped particle under the influence of a time-dependent force as well as friction and external noise. The basic equation of motion for the position $x(t)$ of the particle in the laboratory frame reads
\begin{eqnarray}
\label{oscillator}
\alpha \dot{x}(t)=-\kappa(x(t)-vt)+\xi(t).
\end{eqnarray}
Here, the force $-\kappa(x(t)-vt)$ stems from a parabolic potential $U(x,t)=\kappa(x-vt)^2/2$ which moves with constant velocity $v$, where either $v>0$ or $v<0$. The parameter $\kappa$ denotes the strength of the potential, $\alpha$ the friction coefficient, and $\xi(t)$ stochastic noise from the environment. Similar models, where $\xi(t)$ is given by thermal Gaussian and external L\'evy noise respectively have been investigated in \cite{VanZon04,Touchette07}. If we move the potential for a time period $\tau$, a certain amount of work is done on the particle, namely
\begin{eqnarray}
\label{work}
W_\tau=-\kappa v\int_0^\tau (x(t)-vt) \upd t.
\end{eqnarray}
In the following we are interested in the properties of the steady state work fluctuations in the model Eq.~(\ref{oscillator}), when the noise $\xi(t)$ is given by a sequence of stochastic kicks of variable frequency $\lambda$ and amplitude $\Gamma$. More precisely, we consider Poissonian shot noise of the form \cite{Feynman}
\begin{eqnarray}
\label{F_poiss_noise}
z(t)=\sum_{k=1}^{n_t}\Gamma_k\delta(t-t_k),
\end{eqnarray}  
where $n_t$, the number of kicks in time $t$, is determined by the Poisson counting process $P(n_t)=(\lambda t)^n e^{-\lambda t}/n!$. The parameter $\lambda$ denotes the average number of kicks per unit time (rate of kicks) so that there are $\lambda t$ kicks occurring in the time interval $[0,t]$. When a kick occurs, the amplitude $\Gamma_k$ is sampled randomly from a distribution $p(\Gamma)$. In the following we focus on \textit{one-sided} shot noise and assume an exponential distribution of amplitudes $p(\Gamma)=\Gamma_0^{-1}e^{-\Gamma/\Gamma_0}$. Noise specified according to Eq.~(\ref{F_poiss_noise}) is characterized by delta-correlated cumulants and mean $\left<z(t)\right>=\lambda\Gamma_0$ \cite{Rodriguez85}. It is convenient to use a noise with a zero mean so that we consider in Eq.~(\ref{oscillator})
\begin{eqnarray}
\label{xi}
\xi(t)=z(t)-\lambda\Gamma_0.
\end{eqnarray}

Using this form of the noise and changing to coordinates in a comoving frame, $y(t)\equiv x(t)-vt$, the equation of motion~(\ref{oscillator}) reads
\begin{eqnarray}
\label{F_y}
\dot{y}(t)=-\frac{1}{\tau_r} y(t)-\left(v+\frac{\lambda\Gamma_0}{\alpha}\right)+\frac{1}{\alpha}z(t),
\end{eqnarray}
and the work Eq.~(\ref{work}) is given as
\begin{eqnarray}
\label{work2}
W_\tau=-\kappa v\int_0^\tau y(t)\upd t.
\end{eqnarray}
In Eq.~(\ref{F_y}) $\tau_r\equiv\alpha/\kappa$ denotes the characteristic relaxation time of the oscillator. It is crucial that Poissonian shot noise gives rise to two additional time scales in our model, namely the characteristic time scale of the fluctuations $\tau_\lambda\equiv\lambda^{-1}$, which is the mean waiting time between two successive kicks, and the time scale
\begin{eqnarray}
\label{tau_p}
\tau_p\equiv\frac{\Gamma_0}{\alpha|v|},
\end{eqnarray}
which relates the mean amplitude of fluctuations and the friction due to the driving. We will see below that the three time scales $\tau_r$, $\tau_\lambda$, and $\tau_p$ determine the properties of the work fluctuations in our model, and that critical behavior appears due to the interplay of these times.

From the Langevin equation~(\ref{F_y}) we can infer two important properties of the model. Firstly, upon averaging of Eq.~(\ref{F_y}) we obtain the mean position in the steady state $\left<y(t)\right>=-v\tau_r$ and from Eq.~(\ref{work2}) the mean work $\left<W_\tau\right>=\alpha v^2\tau$, which is always positive in agreement with the second law. Secondly, we find that there exists a \textit{minimal value} $y^*$ of the position coordinate. This can be seen if we solve Eq.~(\ref{F_y}) without the stochastic term $z(t)$, which yields $y^*\equiv-(v+\lambda \Gamma_0/\alpha)\tau_r$ in the long time limit.
The important observation is that the influence of the noise only provides kicks in the positive direction (cf. Eq.~(\ref{F_poiss_noise})) so that $y^*$ is a \textit{cut-off}, i.e., no positions below $y^*$ can be reached. Since the work rate is proportional to the position, we likewise obtain from Eq.~(\ref{work2}) an \textit{extremal value of the work}, namely
\begin{eqnarray}
\label{work_tp}
W^*_\tau=\left<W_\tau\right>(1\pm\lambda\tau_p),
\end{eqnarray}
where the $+$ sign corresponds to $v>0$ and the $-$ sign to $v<0$. It is important to note that the spatial asymmetry of the noise induces a qualitative different behavior of the work fluctuations depending on the sign of $v$. In the case $v>0$ the value of $W_\tau^*$ is always positive and corresponds to the maximum work done on the system in time $\tau$. For $v<0$ on the other hand, $W_\tau^*$ is the minimum work value. In that case $W_\tau^*$ can be either positive or negative (cf. Eq.~(\ref{work2})). Its sign is determined by the two time scales $\tau_\lambda$ and $\tau_p$. From Eq.~(\ref{work_tp}) we see that $W_\tau^*$ is positive if $\tau_\lambda>\tau_p$ and no negative work fluctuations can occur. Furthermore, $W^*_\tau=0$ if $\tau_p=\tau_\lambda$ and $W^*_\tau<0$ if $\tau_\lambda<\tau_p$. The position cut-off $y^*$ represents an infinite barrier in the potential so that the noise induces an \textit{effective non-linearity} in the potential. A work cut-off can also be observed in a Brownian particle model, where the moving potential is given as a non-linear potential of the Lennard-Jones-type \cite{Dykman}. Furthermore, due to these cut-offs, the distributions of position and work are generally non-Gaussian unless one considers the Gaussian limit of the Poissonian shot noise: $\lambda\rightarrow\infty$ and $\Gamma_0\rightarrow 0$ with $\lambda\Gamma_0^2=const$. In this limit $y^*\rightarrow -\infty$ and $W^*_\tau\rightarrow\pm\infty$.

In order to determine the work distribution for arbitrary $\lambda$ and $\Gamma$ values we first derive an exact expression for the characteristic function of the work using a theorem on generalized Ornstein-Uhlenbeck processes \cite{Caceres97,Touchette07}. This theorem states that the characteristic functional of the process $y(t)$ of Eq.~(\ref{F_y}), defined as
\begin{eqnarray}
\label{y_cf}
G_y[h(t)]=\left<\exp\left\{i\int_0^\infty h(t)y(t)\upd t\right\}\right>,
\end{eqnarray}
is determined from the noise functional via
\begin{eqnarray}
G_y[h(t)]=e^{iy_0k_0}G_{\xi'}[k(t)],
\end{eqnarray}
where $G_{\xi'}$ denotes the characteristic functional of the noise plus the drift terms in Eq.~(\ref{F_y}): $\xi'(t)\equiv\xi(t)/\alpha-v$. Furthermore, the functions $k_0$ and $k(t)$ are related to the test function $h(t)$ according to $k(t)=\int_\tau^\infty h(s)e^{(t-s)/\tau_r}\upd s$ and $k_0=\int_0^\infty h(s)e^{-s/\tau_r}\upd s$ \cite{Caceres97}. It is then important to note that choosing the particular test function $\bar{h}(t)=-q\kappa v\Theta(\tau-t)$ in Eq.~(\ref{y_cf}) leads to
\begin{eqnarray}
G_y[\bar{h}(t)]=\left<\exp\left\{-iq \kappa v\int_0^\tau y(t) \upd t\right\}\right>=G_{W_\tau}(q),
\end{eqnarray}
i.e., the characteristic functional for $\bar{h}$ reduces to the characteristic function of the work $G_{W_\tau}(q)\equiv \left<e^{iqW_\tau}\right>$ due to Eq.~(\ref{work2}) \cite{Touchette07}. In our case the noise functional $G_{\xi'}[k(t)]$ is given by
\begin{eqnarray}
\label{F_nG_cb}
G_{\xi'}[k(t)]&=&e^{-i(v+\lambda\Gamma_0/\alpha)\int_0^\infty k(t)\upd t}G_z[k(t)/\alpha],
\end{eqnarray}
where $G_z$ denotes the characteristic functional of the Poissonian shot noise $z(t)$ whose exact analytical expression is known \cite{Feynman}:
\begin{eqnarray}
\label{cf_poisson}
G_z[k(t)]=\exp\left\{\lambda\int_0^\infty\left(\left<e^{i\Gamma k(t)}\right>_\Gamma-1\right)\upd t\right\}.
\end{eqnarray}

Using this noise functional together with the functions $k_0$ and $k(t)$ obtained with $\bar{h}(t)$ yields, after some manipulation, the characteristic work function
\begin{eqnarray}
\label{cf_final}
G_{W_\tau}&\propto&\left(1+iq\Gamma_0v(1-e^{-\tau/\tau_r})\right)^{\lambda\tau_r\left(\frac{1}{1+iq \Gamma_0 v}-1\right)}\nonumber\\
&&\exp\left\{-iq\alpha v y^*\frac{\tau}{\tau_r}+\lambda\tau\left(\frac{1}{1+iq \Gamma_0 v}-1\right)\right\},\nonumber\\
\end{eqnarray}
with a pole at $q=i/(\Gamma_0 \alpha v)$. In the derivation of Eq.~(\ref{cf_final}) we have made use of the exponential distribution of amplitudes $\Gamma$ and furthermore chosen an initial condition $y_0$ sampled from the nonequilibrium steady state distribution, which can be found by solving the Fokker-Planck equation associated with the Langevin equation~(\ref{F_y}) \cite{Rodriguez85}.

For our discussion of the work fluctuations we introduce the scaled dimensionless work value $p$, defined by
$p\equiv W_\tau/\left<W_\tau\right>$. The distribution of $p$ is obtained from the inverse Fourier-transform of $G_{W_\tau}$
\begin{eqnarray}
\label{inverseFT}
\Pi_\tau(p)=N\frac{\alpha v^2\tau}{2\pi}\int_{-\infty}^\infty G_{W_\tau}(q)e^{-iqp\alpha v^2\tau}\upd q,
\end{eqnarray}
where $N$ denotes the normalization constant. To our knowledge there is no exact result for the inverse Fourier transform Eq.~(\ref{inverseFT}). However, for large $\tau$ the integral will be dominated by its saddle-point and can then be analytically obtained using the method of steepest descent \cite{Jeffreys}. Neglecting terms of order $\tau^{-1/2}$ then yields the following result for the distribution $\Pi_\tau(p)$:
\begin{eqnarray}
\label{p_dist}
\Pi_\tau(p)&\cong&\frac{N}{\sqrt{4\pi}}\frac{\sqrt{\tau/\tau_\lambda}}{|p^*-1|}\left(\sqrt{\frac{p^*-p}{p^*-1}}\right)^{-\frac{\tau_r}{\tau_\lambda}\left(\sqrt{\frac{p^*-p}{p^*-1}}-1\right)-\frac{3}{2}}\nonumber\\
&&\times\exp\left\{-\frac{\tau}{\tau_\lambda}\left(\sqrt{\frac{p^*-p}{p^*-1}}-1\right)^2\right\},
\end{eqnarray}
where $p^*$ denotes the rescaled extremal value of work: $p^*\equiv W^*_\tau/\left<W_\tau\right>=1\pm\tau_p/\tau_\lambda$ (cf. Eq.~(\ref{work_tp})). Here, as well as in the rest of this letter, we have expressed $\lambda$ as $\tau_\lambda^{-1}$ in order to emphasize the crucial role of the time scales. Eq.~(\ref{p_dist}) shows that the distribution $\Pi_\tau(p)$ is completely specified by the times $\tau_r$, $\tau_\lambda$, $\tau_p$, and $\tau$. Importantly, the square root $\sqrt{(p^*-p)/(p^*-1)}$ is always real, since both $p^*-p$ and $p^*-1$ are either positive ($v>0$) or negative ($v<0$). For $|p^*-p|\gg 0$ the tail decays faster than exponential but becomes exponential for very large $\tau$, where $\Pi_\tau(p)$ exhibits the large deviation form
\begin{eqnarray}
\label{largeDev}
\Pi_\tau(p)\cong e^{-\tau I(p)}
\end{eqnarray}
with rate function
\begin{eqnarray}
\label{rate_func}
I(p)\equiv\frac{1}{\tau_\lambda}\left(\sqrt{\frac{p^*-p}{p^*-1}}-1\right)^2.
\end{eqnarray}

\begin{figure}
\begin{center}
\includegraphics[height=5cm]{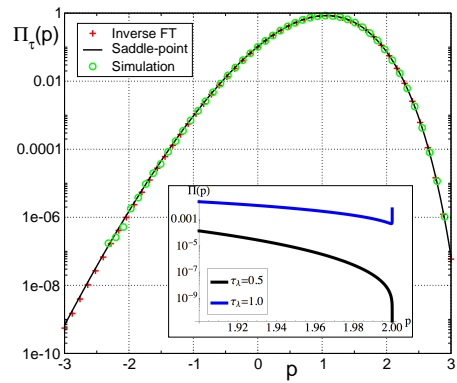}
\caption{\label{Fig_dist}Comparison of the analytic saddle-point approximation Eq.~(\ref{p_dist}) with a numerical inverse Fourier transform of $G_{W_\tau}$ and results from a direct simulation of the equation of motion (\ref{F_y}). Parameter values: $\tau=10\tau_r$, $\tau_r=1$, $v=1$, $\tau_\lambda=0.2$, $\Gamma_0=0.5$. Inset: The approach to the cut-off given by Eq.~(\ref{p_dist}). For $\tau_r/\tau_\lambda>3/2$ the distribution goes to zero (black line), while it diverges when $\tau_r/\tau_\lambda<3/2$ (blue line). Parameter values: $\tau=10$, $\tau_r=1$, $\tau_p=1$.}
\end{center}
\end{figure}

Furthermore, one notices two different singularities appearing in Eq.~(\ref{p_dist}). Firstly, the derivative of $\Pi_\tau(p)$ diverges for $p\rightarrow p^*$ as $\Pi'(p)\propto|p^*-p|^{-1}$ in leading order. This means that the approach of $\Pi_\tau(p)$ to the cut-off has a vertical slope (see inset Fig.~\ref{Fig_dist}). Secondly, one notices that $\Pi_\tau(p)$ itself diverges for $p\rightarrow p^*$ if $\tau_r/\tau_\lambda<3/2$. This divergence for small $\tau_r/\tau_\lambda$ at $p=p^*$ can be understood by considering the nature of the work and position cut-offs in our model. If $\tau_r/\tau_\lambda$ is too small the system relaxes `too quickly' in between the stochastic kicks and thus spends most of its time at the position that it would assume without noise, i.e., at $y^*$. Consequently, the particle will predominantly acquire work $W^*_\tau$ over time $\tau$ leading to a divergence in the work distribution at $p=p^*$ (see inset Fig.~\ref{Fig_dist}). The condition $\tau_r/\tau_\lambda<3/2$ can also be expressed in terms of a \textit{critical friction coefficient} $\alpha^*=\frac{3}{2}\kappa\tau_\lambda$, so that the divergence of $\Pi_\tau(p)$ appears when $\alpha<\alpha^*$.

We find that Eq.~(\ref{p_dist}) yields an excellent approximation of the distribution $\Pi_\tau(p)$ at least for $\tau\geq 10\tau_r$. This becomes evident in Fig.~\ref{Fig_dist}, where we compare Eq.~(\ref{p_dist}) with a numerical inverse Fourier transform of $G_{W_\tau}$ and also with results from a direct simulation of the equation of motion (\ref{F_y}) using a Poissonian increment method \cite{Kim07}.

In the asymptotic regime $\tau\rightarrow \infty$ the work distribution Eq.~(\ref{p_dist}) is dominated by the large deviation form Eq.~(\ref{largeDev}). In order to further discuss the fluctuation properties of work we consider the fluctuation function
\begin{eqnarray}
\label{fluctuations}
f_\tau(p)\equiv \frac{1}{\left<W_\tau\right>}\ln\frac{\Pi_\tau(p)}{\Pi_\tau(-p)}.
\end{eqnarray}
The conventional fluctuation theorem then predicts that $\lim_{\tau\rightarrow\infty}f_\tau(p)=p$. From Eq.~(\ref{largeDev}) we obtain instead
\begin{eqnarray}
\label{fluc_func}
&&\lim_{\tau\rightarrow\infty}f_\tau(p)\nonumber\\
&&=\frac{2p}{v\Gamma_0}+\frac{2(p^*-1)}{v\Gamma_0}\left(\sqrt{\frac{p^*-p}{p^*-1}}-\sqrt{\frac{p^*+p}{p^*-1}}\right),
\end{eqnarray}
defined on the interval $[-p^*,p^*]$. This restriction on the range of $p$-values is similar to deterministic Anosov systems, where the phase-space is bounded \cite{Gallavotti95}. We see that the conventional fluctuation theorem is violated in our model, even though we have identified a large deviation form of the distribution. We characterize the behavior of the work fluctuations for $v>0$ and $v<0$ separately.

\begin{figure}
\begin{center}
\includegraphics[height=5cm]{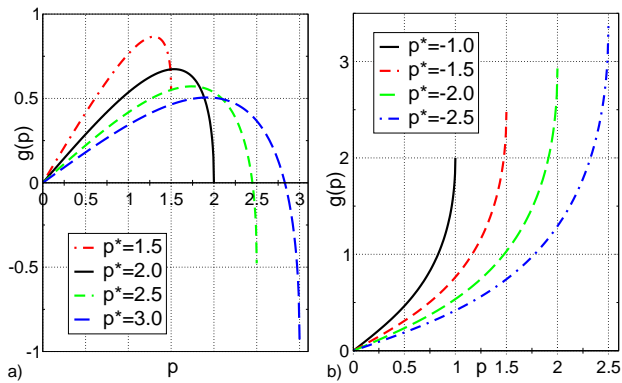}
\caption{\label{Fig_ft}The fluctuation function $g(p)$. At the cut-off $g'(p)$ diverges while $g(p)$ assumes a finite value. a) $v>0$ and four different $p^*$ values. For $p^*>2$ we observe that $g(p)$ becomes negative in the interval $p\in[p_0,p^*]$. b) $v<0$ and four different $p^*$ values. Here, $g(p)$ only has a zero at $p=0$. }
\end{center}
\end{figure}

(i) For $v>0$ we find that Eq.~(\ref{fluc_func}) has the zeros $p_1=0$ and $p_2=2\sqrt{p^*-1}$ (due to the symmetry of $f_\tau(p)$ we neglect the negative root). The zero $p_2$ becomes significant when $p^*>2$, because then $p^*>p_2$ and $p$ can assume values in the interval $[p_2,p^*]$. The crucial observation is that $\lim_{\tau\rightarrow\infty}f_\tau(p)$ from Eq.~(\ref{fluc_func}) becomes negative for $p\in[p_2,p^*]$, if $p^*>2$. This is evident in Fig.~\ref{Fig_ft}a), where we plot the dimensionless rescaled fluctuation function $g(p)\equiv\lim_{\tau\rightarrow\infty}|v|\Gamma_0f_\tau(p)$, which depends only on $p^*$. There exists therefore a parameter regime in which negative fluctuations of a certain magnitude are \textit{more likely to occur} than corresponding positive ones. In fact, since $p^*=1+\tau_p/\tau_\lambda$ we find that $p^*>2$ if $\tau_p>\tau_\lambda$. Using the definition of $\tau_p$, Eq.~(\ref{tau_p}), we can identify a corresponding \textit{critical velocity} $v^*\equiv\Gamma_0/(\alpha\tau_\lambda)$ so that for $0<v<v^*$ the fluctuation function is negative.

This surprising property originates from the strongly asymmetric tails of the work distribution $\Pi_\tau(p)$: The negative tail decays exponentially while the positive tail is bounded by the cut-off at $p^*$. It is important to note that despite the existence of a negative regime of the fluctuation function $f_\tau(p)$, the second law is never violated: the mean value of the work is always positive, $\left<W_\tau\right>=\alpha v^2\tau$. For $p\rightarrow p^*$ the derivative $g'(p)$ diverges like $(p^*-p)^{-1/2}$, while $g(p)$ remains finite (see Fig.~\ref{Fig_ft}a)).

(ii) For $v<0$ we have to distinguish two regimes. Firstly, for $\tau_\lambda>\tau_p$ no negative work fluctuations can occur (cf. Eq.~(\ref{work_tp})) and therefore the fluctuation function $f_\tau(p)$ can not be defined in this parameter regime. Secondly, for $\tau_\lambda<\tau_p$ we have $p^*<0$ and we can discuss $g(p)$ in the interval $[0,-p^*]$. In this case $g(p)$ is only zero at $p=0$. As in case (i), the derivative $g'(p)$ diverges for $p\rightarrow p^*$ while $g(p)$ remains finite (see Fig.~\ref{Fig_ft}b).

We note that our theory could be adapted to an experiment similar to that of Mahadevan \textit{et al} \cite{Mahadevan03}, where a lubricated rod of a hydrogel sliding on a soft vibrating substrate is considered as a model for biomimetic ratcheting motion. Instead of the purely oscillatory vibrations of \cite{Mahadevan03} one could induce asymmetric Poissonian shot noise, which could lead to work fluctuations with similar features as those presented in this letter.

In summary, we have investigated the work fluctuations of a particle, confined to a moving harmonic potential, under the influence of friction and Poissonian shot noise. The one-sidedness of the fluctuations leads to a cut-off in the work distribution, which is therefore strongly non-Gaussian. We have shown that the work distribution exhibits a large deviation form but that, nevertheless, the conventional fluctuation theorem is violated.

The authors thank Dr. Hugo Touchette for stimulating discussions. They also gratefully acknowledge financial support of the National Science Foundation under award PHY-0501315  and of the EPSRC, grant no. GR/T24593/01.

\end{document}